 \documentclass[preprint,onecolumn,showpacs,preprintnumbers,amsmath,amssymb,floatfix]{revtex4}
\usepackage{graphicx}
\usepackage{subfigure}
\usepackage{dcolumn}
\usepackage{bm}
\usepackage{verbatim}

\begin{document}
\title{An exact method to compute a $p$-value for the beyond-pairwise correlations among cancer gene mutations}
\author{Jaegil Kim}
\email{jaegil@broadinstitute.org}
\affiliation{Broad Institute of MIT and Harvard,  Cambridge, MA 02142}
\author{Atanas Kamburov}
\affiliation{Broad Institute of MIT and Harvard,  Cambridge, MA 02142}
\author{Michal Lawrence}
\affiliation{Broad Institute of MIT and Harvard,  Cambridge, MA 02142}
\author{Yosef Maruvka}
\affiliation{Broad Institute of MIT and Harvard,  Cambridge, MA 02142}
\author{Gad Getz}
\email{gadgetz@broadinstitute.org}
\affiliation{Broad Institute of MIT and Harvard,  Cambridge, MA 02142}



\begin{abstract}
The increasing observation of mutual exclusivity correlations among cancer gene mutations is a key component for identifying driver events or pathways in cancer genome analysis. 
Here we report a rigorous statistical method to compute an exact $p$-value for the beyond-pairwise mutual exclusivity or co-occurrence relationships among cancer gene mutations by enumerating a null distribution of overlapping mutations across more than two genes. The validity and the advantage of our method is explicitly demonstrated in both cancer gene mutations and simulation data through the comparison to the permutation test. 
\end{abstract}
\maketitle

\section{Introduction}
The recent development of high-throughput genome sequencing technologies reveals a remarkable complexity in genetic and epigenetic aberrations characteristic of cancer initiation and progression~\citep{Jones,Hanahan}. Given genetic heterogeneities across samples even in a single cancer type one of challenges is to distinguish underlying driver events from a myriad of random passenger events~\citep{Vogelstein,Getz}. The standard approach for discovering cancer driver genes is to  identify genes with a significantly higher mutation rate across samples than the background mutation rate~\citep{Mike}. In addition to the mutational recurrence across samples the recent cancer gene discovery  is often corroborated by evaluating a significance of individual mutations or genes in the context of mutation patterns projected onto cellular signaling and regulatory pathways~\citep{Ciriello2}. The mutual exclusivity relationship is commonly observed among driver mutations on the same pathway and plays an important role in discovering cancer driver events or pathways in many network and pathway analysis algorithms~\citep{Ciriello1,Vandin,Miller}.
 
Fisher's exact method~\citep{Fisher} or hypergeometric test provides an exact $p$-value for the pairwise mutual exclusivity in two genes by directly enumerating more extreme cases with overlapping mutations equal or less than the observed one. However, it is non-trivial to extend the pairwise exact test to the gene set consisting of more than two genes. Considering an illuminating role of mutation patterns in the discovery of cancer driver genes or pathways it is very demanding to develop a systematic method to evaluate a statistical significance of the beyond-pairwise correlations or anti-correlation among mutations. In this paper, we report a rigorous statistical method to determine an exact  $p$-value for the beyond-pairwise mutual exclusivity or co-occurrence relationships among cancer mutations in an arbitrary number of genes. Our study illustrates that a null distribution of overlapping mutations across genes can be determined via a sequential operation of a \textit{hypergeometric sampling} and the exact $p$-value can be computed through a simple recursive formula for given mutation data. We validated our method in both cancer gene mutations and simulation data through the comparison to the permutation test. 

\section{Methods}
Let us start by briefly reviewing the pairwise mutual exclusivity, in which two genes $g_1$ and $g_2$ had $n_1$ and $n_2$ mutations, respectively, across $n$ samples. When $n_1$ mutations in $g_1$ are randomly chosen from $n$ samples, and another random samples lead to $n_2$ mutations in $g_2$, the probability that two genes have $z$ overlapping mutations exactly follows a hypergeometric distribution for $z\in [\textrm{max}(0,n_1+n_2-n), \textrm{min}(n_1,n_2)]$ as 
\begin{equation}
P_2(z)= H(z; n_1,n-n_1,n_2)= \frac{\binom{n_1}{z} \binom{n-n_1}{n_2-z}}{\binom{n}{n_2}},
\label{hyper}
\end{equation}
$\binom{n}{m}$ being a binomial coefficient. 
The distribution $P_2(z)$ describes a \textit{hypergeometric sampling} characterized by the probability of choosing $z$ white balls in $n_2$ draws without replacements from a finite population of $n_1$ white balls and $(n-n_1)$ black balls. 
The $p$-value for the pairwise exclusivity is determined by summing all probabilities of more extreme cases with $z \le z_0$, $z_0$ being the observed overlapping mutations, as 
$ \sum_{z=z^*}^{z_0} P_2(z)$, $z^*=\textrm{max}(0,n_1+n_2-n)$.
The hypergeometric test with Eq.~(\ref{hyper}) is identical to the one tailed version of Fisher's exact test.


The mutation data for $m$ genes with each $n_i$ mutations ($i=1,\ldots,m$) across $n$ samples can be represented by $n\times m$ binary matrix $A$, $A_{ij}=1$ if gene $g_j$ is mutated in sample $i$, and $0$ otherwise. Here we quantified the overlapping mutations by $z=\Gamma(A)-\omega(A)$, $\Gamma(A)= \sum_{ij} A_{ij}$ and $\omega(A)= \sum_i \textrm{max}(A_{i1},\ldots,A_{im})$. Note that $\omega(A)$ represents  the number of samples having at least one mutation across genes and $z$ ($\ge 0$) becomes zero only if mutations among genes are perfectly exclusive.

Our goal is to construct a random null distribution characterized by the probability that the set of m genes ($m \ge 3$) have $z$ overlapping mutations. 
For this purpose we first consider the simplest case of $m=3$. Without a loss of generality we  assume $n_1 \ge n_2 \ge n_3$ by re-ordering genes. Since the probability for $x$ overlapping mutations between $g_1$ and $g_2$ is already determined as $H(x; n_1,n-n_1,n_2)$ the remaining step is to enumerate newly formed overlapping mutations with an addition of $g_3$ for the fixed overlaps $x$ between $g_1$ and $g_2$. The key finding is that the probability that all three genes have extra $y$ overlapping mutations in addition to $x$ is also determined by a \textit{hypergeometric sampling}, yielding $H(y; p, q, n_3)$ for $y \in [\textrm{max}(0,p+n_3-n), \textrm{min}(p,n_3)]$, $p=\textrm{max}(n_1+n_2-x,n)$ and $q=n-p$. In analogous to the pairwise mutual exclusivity, this distribution describes a likelihood of choosing $y$ white balls in $n_3$ draws without replacements from a finite population of $n$ size comprised of $p$ white balls and $q$ black balls. Note that the number of white balls is now $(n_1+n_2-x)$ due to existing $x$ overlapping mutations between  $g_1$ and $g_2$. 

Summing probabilities for all possible choices of $x$ and $y$ the final null distribution of observing $z$ overlapping mutations is obtained as
\begin{equation}
P_3(z) =  \sum_{x=x^*}^{n_2} \sum_{y=y^*}^{n_3} P_2(x)  H(y;p,q,n_3)   \delta\left[z-(x+y)\right]
\label{P3}
\end{equation}
$x^*=\textrm{max}(0,n_1+n_2-n)$, $y^*=\textrm{max}(0,n_1+n_2+n_3-n)$, $p=\textrm{max}(n_1+n_2-x,n)$, and $q=n-p$. For an arbitrary $m$ ($\ge 3)$ $P_m(z)$ is iteratively determined using a following recursive relation,
\begin{equation}
P_m(z)= \sum_{x=x^*}^{{n}_{m}^*} \sum_{y=y^*}^{n_m} P_{m-1}(x) H(y; p,q,n_m) \delta\left[z-(x+y)\right]
\label{Pm}
\end{equation}
where $x^*=\textrm{max}(0,n_1+n_m^*-n)$, $y^*=\textrm{max}(0,n_1+n_m^*+n_m)$, $p= \textrm{max}(n_1+n_m^*-x,n)$ and $q= n-p$. Here ${n}_m^{*}= \sum_{j=2}^{m-1} n_j$ denotes the number of maximum overlapping mutations in the gene set consisting of $g_1,\ldots,g_{m-1}$.

\section{Results}

\begin{figure}
\scalebox{0.65}{\includegraphics[bb= 0 0 800 125,clip]{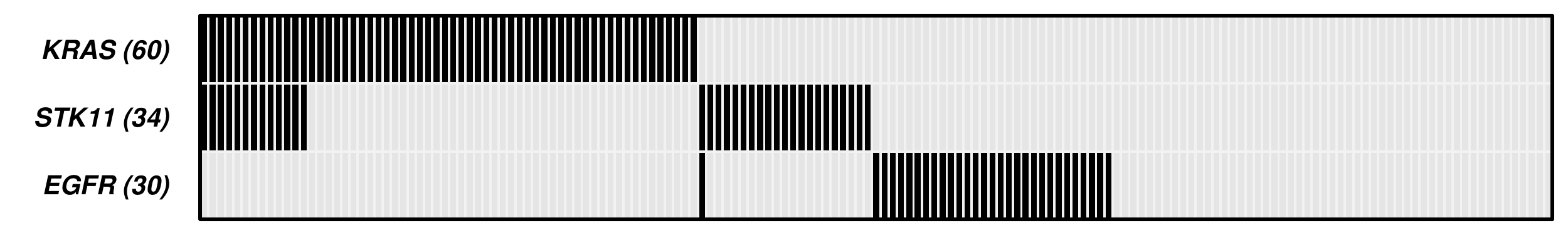}}
\caption{Mutations in {\it KRAS}, {\it STK11}, and {\it EGFR} in lung adenocarcinoma project~\citep{Ding}}
\label{Fig0}
\end{figure}

\begin{figure}
\scalebox{0.525}{\includegraphics[bb= 0 0 500 580,clip]{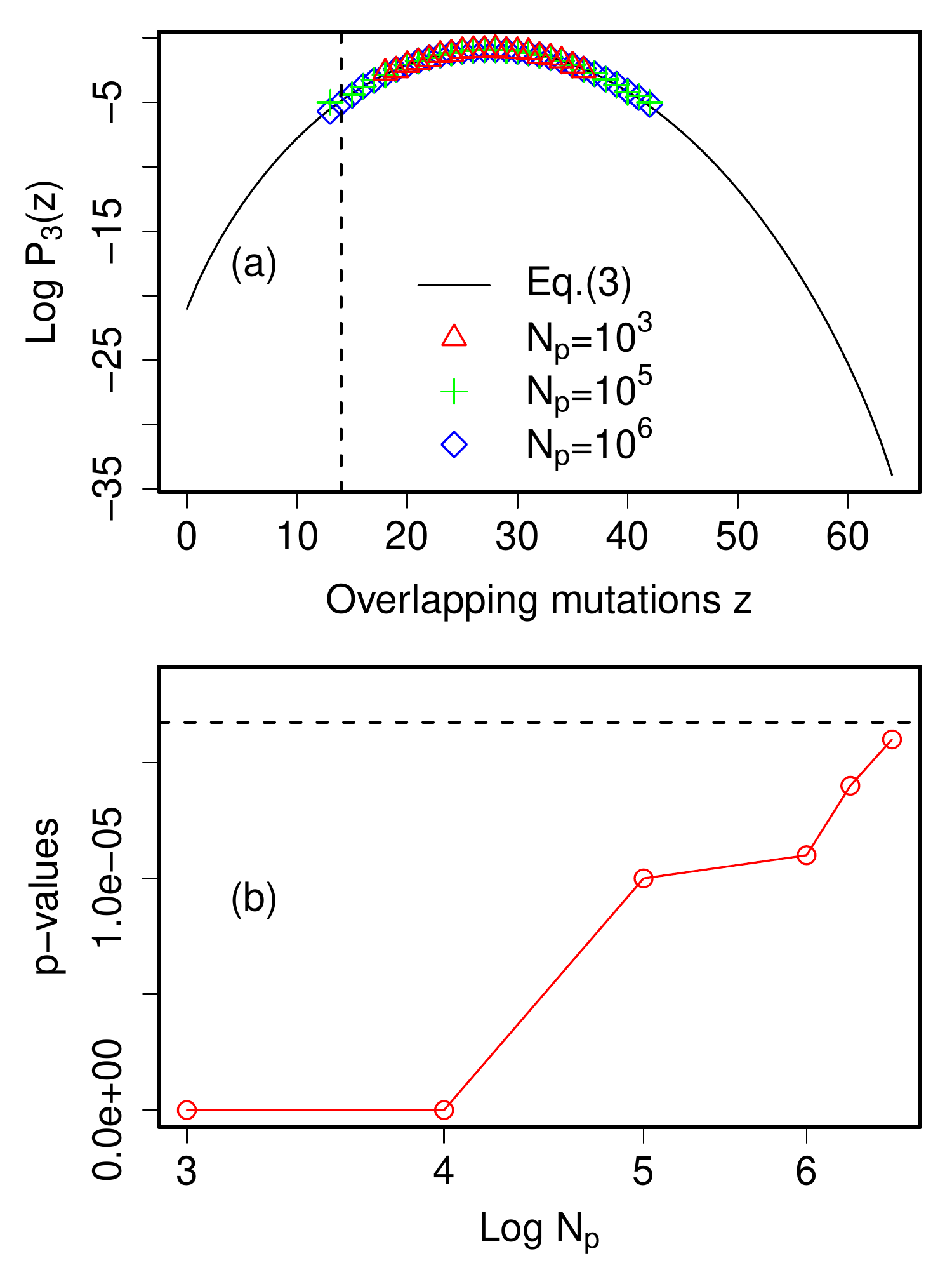}}
\caption{(a) The exact null distribution $\textrm{Log}[P_3(z)]$ of Eq.~(\ref{Pm}) and approximate $\textrm{Log}[\tilde{P}_3(z,N_p)]$ from the permutation test with varying $N_p$ as a function of overlapping mutations $z$ for the gene set, {\it KRAS}, {\it STK11}, and {\it EGFR}, in lung adenocarcinoma project~\citep{Ding}, and (b) $p$-values as a function of $N_p$.  The vertical and horizontal dashed lines corresponds to $z_0$ and the exact $p$-value, respectively.}
\label{Fig1}
\end{figure}

We validated our method in the mutation data of three driver genes, {\it KRAS}, {\it STK11}, and {\it EGFR}, in the lung adenocarcinoma sequencing project~\citep{Ding}. This gene set was identified as the most exclusive one for $m=3$ by DENDRIX analysis~\citep{Vandin}. The mutation data across $163$ samples is represented in Fig.~\ref{Fig0}, resulting in $n_1=60$, $n_2=34$, $n_3=30$, and $z_0=14$. In Fig.~\ref{Fig1}(a), the exact null distribution $P_3(z)$ from Eq.~(\ref{Pm}) is compared with the approximate $\tilde{P}_3(z;N_p)$ determined from the permutation test with varying $N_p$, $N_p$ being the number of permutations. Note that $\tilde{P}_3(z,N_p)$ is only non-zero in a narrow dynamic region $z$ corresponding to $[z_l,z_u]$, $P_3(z_{l,u}) \sim 1/N_p$, while $P_3(z)$ spans a full dynamic range of $z\in [\textrm{max}(0,\sum_i^m n_i - n), \textrm{max}(0,\sum_{i=2}^m n_i)]=[0,64]$. The $p$-value from the permutation test in Fig.~\ref{Fig1}(b) asymptotically converges to the exact $p=1.673\times 10^{-5}$ with increasing $N_p$, but $p$-values are zero for small $N_p$ since the exact $p$-value $ < 1/N_p$. We also tested our method in a more complicated case with $n=500$ and $m=5$ using a simulation data (see Supplementary Figure and text for details). Finally, we would like to stress that our method is equally applicable to the significance test of co-occurring mutations by summing probabilities of more extreme cases with $z \ge z_0$, yielding $\textrm{$p$-value}= \sum_{z \ge z_0} P_m(z)$. 

\section{Summary}
In summary, a rigorous statistical method to evaluate a significance of beyond-pairwise mutual correlations among mutation data is developed by establishing an exact null distribution of overlapping mutations across genes. In contrast to the permutation test, which is approximate and computationally demanding with increasing $m$, our method enables an instant calculation of exact $p$-values regardless of the size of interesting gene set, making our method well-suited to an extensive search or prioritizing driver events or pathways in heterogeneous mutation data combined with other network or pathway analysis algorithms.

\section*{Acknowledgement}
\bibliographystyle{bioinformatics}

\newpage

\appendix
\section*{Supplementary Material for "An exact method to compute a  p-value for the beyond-pairwise correlations among cancer gene mutations"}
\section*{Simulation Data}

To validate our method in a more complicated situation we created a following simulated mutation data with $n=500$ and $m=5$ characterized by 
\begin{eqnarray}
A_{ij} &=& \left\{
	\begin{array} {ll}
		1 & \mbox{if $i \in [1,200]$  for $j=1$},  \\
		1 & \mbox{if $i \in [101,250]$ for $j=2$},  \\
		1 & \mbox{if $i \in [201,300]$ for $j=3$}, \\
		1 & \mbox{if $i \in [351,450]$ for $j=4$},  \\
		1 & \mbox{if $i \in [431,480]$ for $j=5$},  \\
		0 & \mbox{otherwise},
	\end{array} 
	\right.
\label{A5_single}
\end{eqnarray}
$n_1=200$, $n_2=150$, $n_3=100$, $n_4=100$, $n_5=50$, and $z_0=170$. 
In Eq.~(\ref{A5_single}), non-zero elements of $A_{ij}$ for each $j$  could be assigned to any samples under the constraints of $n_i$ and the number of overlapping mutations in each pair of genes, and the final null distribution and corresponding $p$-value are not affected by those permutations.   

\begin{figure}
\scalebox{0.45}{\includegraphics[bb= -45 0 550 880,clip]{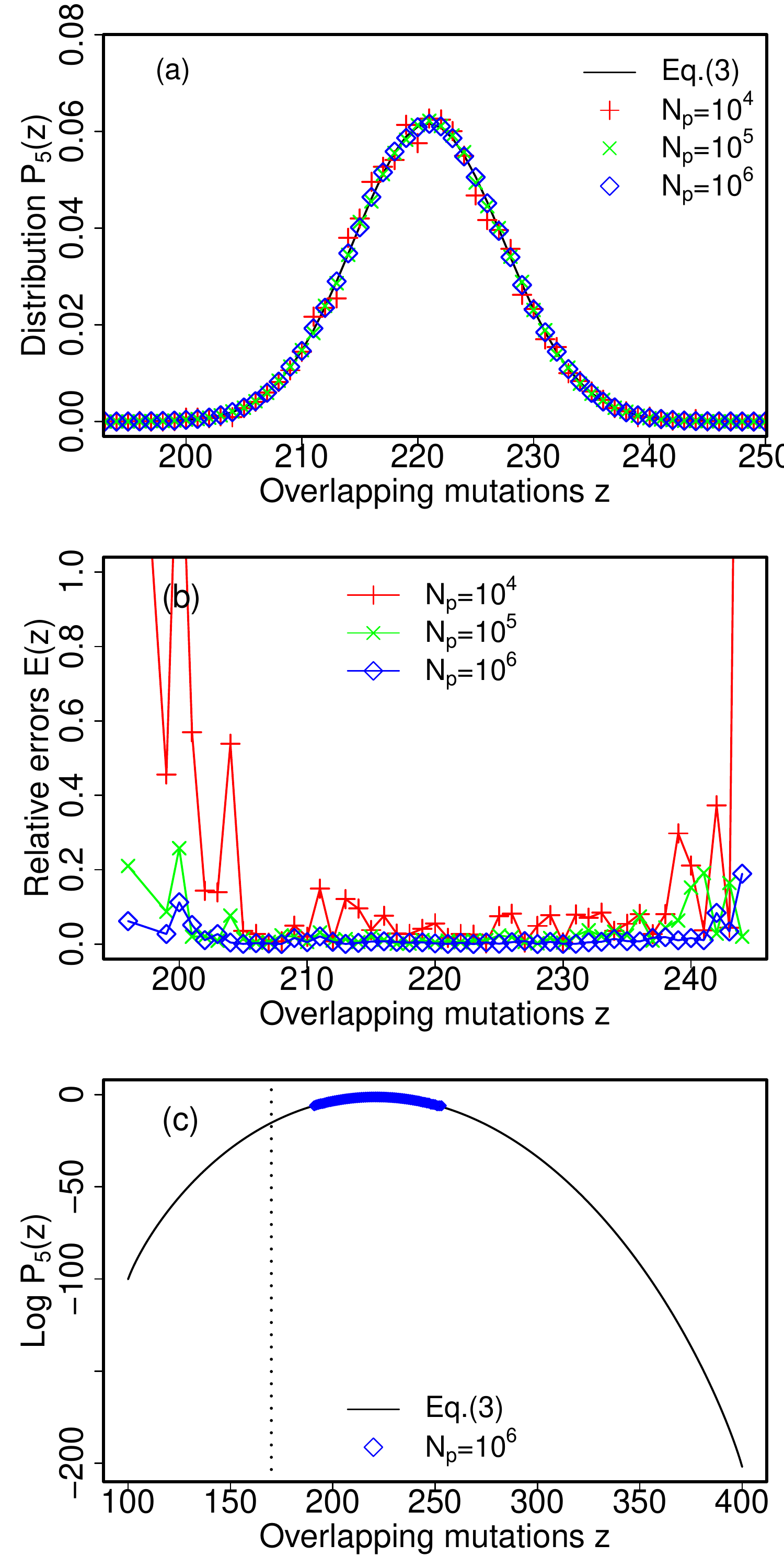}}
\caption{(a) The exact null distribution $P_5(z)$ and approximate distributions $\tilde{P}_5(z,N_p)$ from the permutation test with varying $N_p$ from $10^4$ to $10^6$ as a function of overlapping mutations $z$ for the simulation data in Eq.~(\ref{A5_single}) with $n=500$ and $m=5$, (b) relative errors $E(z)=|(P_5(z)-\tilde{P}_5(z,N_p))/P_5(z)|$ at non-zero $\tilde{P}_5(z,N_p)$, and (c) $\textrm{Log}[P_5(z)]$ (solid line) and $\textrm{Log}[\tilde{P}_5(z,10^6)]$ (diamonds).  Only non-zero values of $\tilde{P}_5(z,10^6)$  were shown and the dashed line corresponds to $z_0$ in (c).}
\label{Fig2}
\end{figure}

The exact null distribution $P_5(z)$ was compared in Fig.~\ref{Fig2}(a) with approximate ones $\tilde{P}_5(z,N_p)$ at various $N_p$, indicating that more than $10^6$ random permutations are needed to provide a fair comparison around $z \sim \bar{z}$, $\bar{z}$ being the mean overlapping mutation, as verified in relative errors in Fig.~\ref{Fig2}(b). However, as clearly seen in Fig.~\ref{Fig2}(b), $\tilde{P}_5(z,N_p)$ provides approximate probabilities in a narrow dynamic region of $z\in [z_l,z_u]$, $P_5(z_{l,u}) \sim 1/N_p$, in comparison to $P_5(z)$ spanning a full dynamic range of $z\in [100,400]$. As expected, the permutation test failed to provide any significance measure for the exclusivity since the correct $p$-value of $8.134\times 10^{-16}$ is far less than the minimum resolution $1/N_p$ in the permutation test. 

\end{document}